\documentclass[aps,prb,amsmath,amssymb,reprint,superscriptaddress]{revtex4-2}

\usepackage[colorlinks,citecolor=blue,linkcolor=red,urlcolor=blue]{hyperref}
\usepackage{color}
\usepackage{graphicx}
\usepackage{subfigure}
\usepackage{verbatim}
\usepackage[normalem]{ulem}
\usepackage{bbm}

\usepackage{physics,amsmath,amsfonts,amssymb,stmaryrd,bm,bbm,bbold}

\begin{document}

\preprint{APS/123-QED}

\title{Casimir-Lifshitz force for graphene-covered gratings}

\author{Youssef Jeyar}
\email{youssef.jeyar@umontpellier.fr}
\affiliation{Laboratoire Charles Coulomb (L2C), UMR 5221 CNRS-{Universit\'e de} Montpellier, F-34095 Montpellier, France}
\author{Minggang Luo}
\affiliation{Laboratoire Charles Coulomb (L2C), UMR 5221 CNRS-{Universit\'e de} Montpellier, F-34095 Montpellier, France}
\author{Brahim Guizal}
\affiliation{Laboratoire Charles Coulomb (L2C), UMR 5221 CNRS-{Universit\'e de} Montpellier, F-34095 Montpellier, France}
\author{H. B. Chan}
\affiliation{Department of Physics, The Hong Kong University of Science and Technology, Clear Water Bay, Kowloon, Hong Kong, China}
\affiliation{William Mong Institute of
Nano Science and Technology, The Hong Kong University of Science and Technology, Clear Water Bay, Kowloon, Hong Kong, China}
\affiliation{Center for
Metamaterial Research, The Hong Kong University of Science and Technology, Clear Water Bay, Kowloon, Hong Kong, China}
\author{Mauro Antezza}
 \email{mauro.antezza@umontpellier.fr}
\affiliation{Laboratoire Charles Coulomb (L2C), UMR 5221 CNRS-{Universit\'e de} Montpellier, F-34095 Montpellier, France}
\affiliation{Institut Universitaire de France, 1 rue Descartes, Paris Cedex 05, F-75231, France}
\date{\today}

\begin{abstract}

We study the Casimir-Lifshitz force (CLF) between a gold plate and a graphene-covered dielectric grating. Using a scattering matrix (S-matrix) approach derived from the Fourier Modal Method (FMM), we find a significant enhancement in the CLF as compared to a mere dielectric slab coated with graphene, over a wide range of temperatures. Additionally, we demonstrate that the CLF depends strongly on the chemical potential of graphene, with maximal effects observed at lower filling fractions. Finally, we analyse the Casimir force gradient between a gold sphere and a graphene-coated dielectric grating, highlighting potential avenues for experimental measurements.

\end{abstract}


\maketitle

\section{\label{sec:Intro}Introduction}

The Casimir-Lifshitz force (CLF), resulting from quantum vacuum fluctuations, manifests in the interactions between polarizable bodies. In 1948, Casimir \cite{casimir_48} first developed a  theoretical framework for this force between two perfectly conducting plates at zero temperature. Later, Lifshitz expanded this framework to encompass bodies with diverse optical properties and at finite temperatures \cite{lifshitz_61}. Since its experimental verification \cite{dalvit2011casimir}, the CLF has attracted significant research interest, with investigations focused on exploring different geometries and materials \cite{Lamoreaux1997prl,Bressi2002prl,Mauro2020prl,Decca2003prl,Antoine2009prl,Nunes2021universe,Schoger2022prl,Bimonte2018prd}.

Among diverse geometrical systems, the grating configuration is particularly important in the study of the CLF and has been explored both theoretically \cite{Lambrecht2008prl,Noto2014pra_dielectirc} and experimentally \cite{PFA_gradF, Nature_2021}, in terms of the non-trivial interplay between geometry and dielectric properties of the interacting bodies. Analyzing such structures necessitates precise and efficient methods such as the Fourier Modal Method (FMM) or its improved version, the FMM with adaptive spacial resolution (FMM-ASR), which significantly enhances efficiency for metallic gratings \cite{Messina2017prb}.

The dielectric properties of the interacting objects can affect the scattering details, which are relevant to both CLF in and out of thermal equilibrium and radiative heat transfer. In particular, there has been much progress in graphene-based structures \cite{PhysRevA.92.062504,jeyarprb2023,Antezza:2023uae,PhysRevB.109.125105,Svetovoy2012prb,Ognjen2012prb,Zheng2017,Volokitin2017Dey,PhysRevLett.126.206802,Chahine2017prl,Liu2021prb,Pablo2022tdm}, that utilizes the unique optical properties of this material \cite{Falkovsky_Graphene_1,Falkovsky_Graphene_2,Awan_Graphene_3,PhysRevB.103.125421,rodriguez2024graphene}. A recent study \cite{PhysRevA_Jeyar} delved into the CLF between graphene gratings, utilizing another enhanced version of the FMM approach known as FMM with local basis functions \cite{Youssef2023pre}. This study shows the potential for modulating the force by manipulating geometry parameters and/or graphene properties. The investigation is further extended to analyzing heat transfer within these structures for both aligned \cite{10.1063/5.0182725} and misaligned \cite{luo2024nearfield} cases. Patterning the graphene sheet into grating will result in a topology transition of the modes for energy transfer from the circular one to the hyperbolic one, which allows a significant enhancement in the heat transfer. 

In another example \cite{10.1063/5.0202512}, graphene is integrated with one-dimensional plasmonic nanogratings with high aspect ratios and narrow grooves, the good agreement between theory and experiment suggests promising prospects for high-performance electrically tunable graphene-based infrared photodetectors and biological sensors.

In this work, we explore the CLF between graphene coated dielectric grating and gold semi-finite plane, using the scattering matrix approach obtained by an adapted FMM to include the graphene sheet. This configuration combines the features of both the grating structures and the special dielectric properties of graphene, which is expected to provide more freedom in controlling the CLF.

In section \ref{sec:Phys_sys}, we describe the physical system, in section \ref{reflexion_operators} we introduce the reflection operators obtained by the FMM. Finally, in section \ref{results}, we present and discuss the numerical results. The calculation details of the scattering matrix are provided in the appendix.

\begin{figure} [htbp]
\vspace{1em}
\centerline {\includegraphics[width=0.4\textwidth]{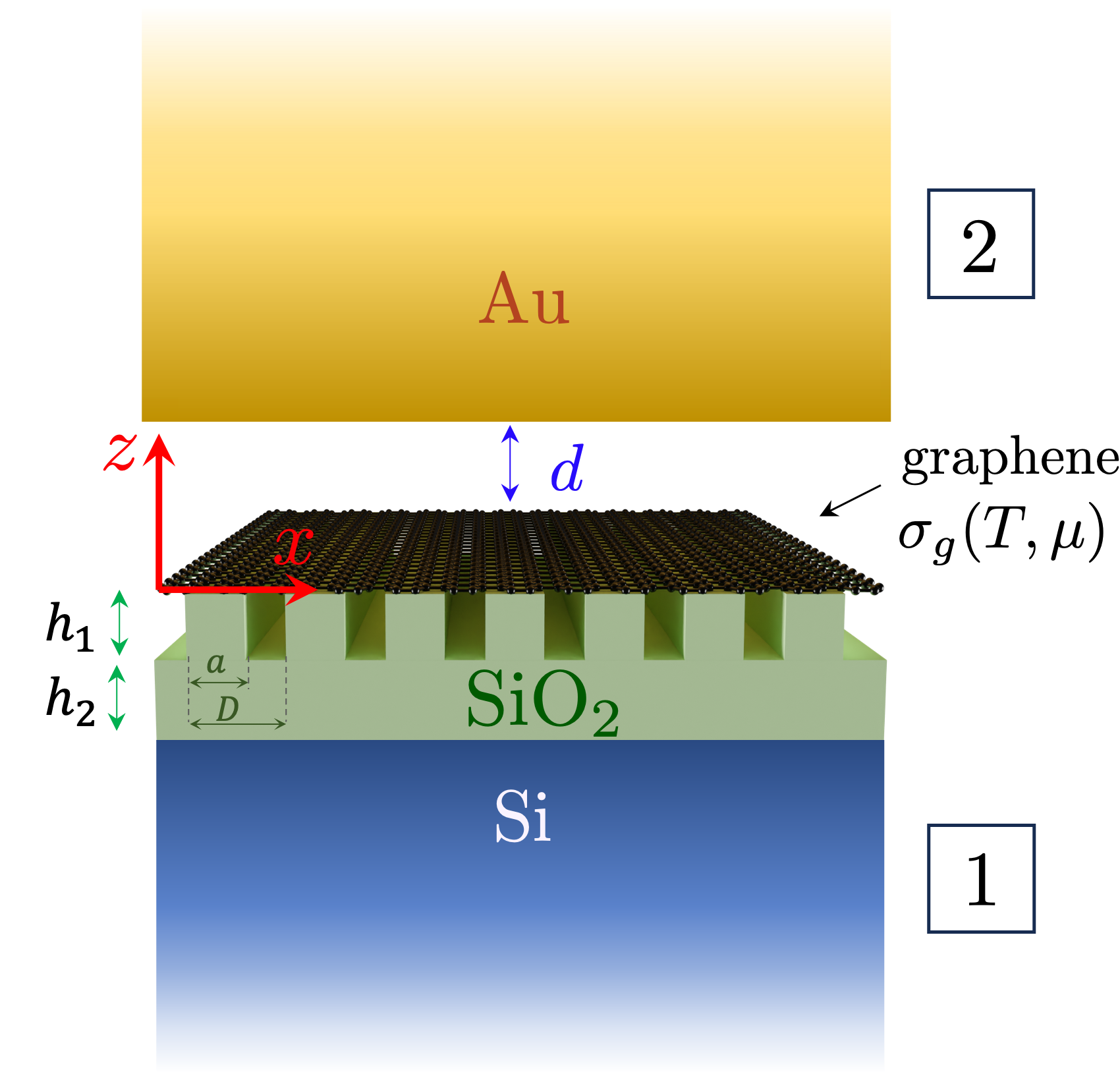}}
\caption{Schematic of the system: a dielectric grating covered with graphene sheet and is placed at a distance $d$ from a gold half-space. The grating is periodic in the $x$ direction with period $D$, characterized by width  $a$, a depth $h_1$ and filling fraction $f = a/D$. }
\label{Physical_system}
\end{figure}

\section{\label{sec:Phys_sys}Physical system}
The physical system depicted in Fig. \ref{Physical_system} consists of a dielectric grating coated with a graphene sheet (body 1) facing a gold half-space (body 2). These two bodies are separated by a distance $d$ and maintained at a temperature $T$. The dielectric grating, composed of fused silica (SiO$_2$), exhibits periodicity in the $x$ direction with a period $D$ (we use a fixed value of $D=1\mu$m in this study), each period contains an SiO$_2$ region of width $a$, giving a filling fraction $f=a/D$. The grating has a thickness $h_1$ of 300 nm and is positioned on top of an SiO$_2$ slab with a thickness $h_2$ of 100 nm, which in turn sits on an silicon ($\text{Si}$) half-space.

For this configuration, the Casimir-Lifshitz pressure (CLP) acting on body 1 along the $z$ positive direction can be expressed as  \cite{Noto2014pra_dielectirc,PhysRevA.84.042102,PhysRevA_Jeyar}
\begin{equation}
P(d,T,\mu)=\frac{ k_{\rm B} T}{4 \pi^2} \sum_{m=0}^{+\infty}{}'  \int_{-\frac{\pi}{D}}^{\frac{\pi}{D}} {\rm d}k_{x}  \int_{-\infty}^{+\infty} {\rm d} k_y   {\rm Tr}\left( \gamma '  \mathcal{M}  \right),
\label{CLP}
\end{equation}
where $\gamma'={\rm diag} (k_{zn}',k_{zn}')_n$, $k_{zn}'=\sqrt{\xi_m^2/c^2+{\bf{k}}_{n}^2}$, ${\bf{k}}_{n}=(k_{xn},k_y)$, $k_{xn}=k_x+n \frac{2\pi}{D}$, $k_{x}$ is in the first Brillouin zone $[-\frac{\pi}{D},\frac{\pi}{D}]$ and $k_y$ is in $\mathbb{R}$. The notation $\operatorname{diag}\left(u_{n},u_{n}\right)_{n}$ represents a diagonal matrix of size $2(2 N+1) \times2(2 N+1)$, where the diagonal elements are $u_{-N}, u_{-N+1}, \ldots, u_{N},u_{-N}, u_{-N+1}, \ldots, u_{N} $. The expression for $\mathcal{M}$ is given by
\begin{equation}
\mathcal{M}= (U^{(12)}\mathcal{R}^{(1)+} \mathcal{R}^{(2)-}+U^{(21)}\mathcal{R}^{(2)-} \mathcal{R}^{(1)+}).
\label{M_operator}
\end{equation}
where $\mathcal{R}^{(1)+}$ and $\mathcal{R}^{(2)-}$ represent the reflection operators corresponding to the two bodies in the (TE, TM) basis obtained using the scattering matrix approach, as will be discussed hereafter. As for the inter-cavity operators, they are defined as follows ($\mathcal{I}$ denoting the operator identity):
\begin{equation}
\begin{aligned}
U^{(12)}&=(\mathcal{I}-\mathcal{R}^{(1)+}\mathcal{R}^{(2)-})^{-1},\\
U^{(21)}&=(\mathcal{I}-\mathcal{R}^{(2)-}\mathcal{R}^{(1)+})^{-1}. 
\end{aligned}
\label{operators}
\end{equation}

In {Eq.} \eqref{CLP}, the sum is over the Matsubara frequencies ($\xi_m=2\pi m k_{\rm B} T/ \hbar$) and the prime on this sum means that the first term ($m$=0) is devided by 2. {$k_{\rm{B}}$ is the Boltzmann constant, and $\hbar$ is the reduced Planck's constant.} 

For fused silica, the dielectric response at imaginary frequencies, is obtained through the Kramers-Kronig relation : $\varepsilon(i\xi_m)=1+2\pi^{-1}\int_{0}^{\infty} \omega\varepsilon''(\omega)/(\omega^2+\xi_m^2)d\omega$ and relies on the data extracted from the dielectric function of SiO$_2$, as sourced from \cite{Book_SiO2}, extrapolated along the real frequency axis.

We adopt the following model for the dielectric function of silicon \cite{Nature_2021}:
\begin{equation}
\varepsilon_{\rm Si}(i\xi_m)= 1.035 + \frac{11.87-1.035}{1+\xi_m^2/\omega_0^2}+\frac{\omega_p^2}{\xi_m(\xi_m+\Gamma)}
\end{equation}
where $\omega_0= 6.6 \times 10^{15}$ rad s$^{-1}$, $\omega_p= 2.37 \times 10^{14}$ rad s$^{-1}$, and $\Gamma= 6.45 \times 10^{13}$ rad s$^{-1}$.

We employ the Drude permittivity model for the dielectric function of gold (Au)
\begin{equation}
\varepsilon_{\rm Au}(i\xi_m)= 1 + \frac{\omega_p^{'2}}{\xi_m(\xi_m+\gamma)}
\end{equation}
with $\hbar \omega_p^{'}$= 9 eV and $\hbar\gamma$= 35 meV.

The contribution of graphene is introduced through its conductivity, which varies with temperature $T$ and chemical potential $\mu$. We model the conductivity as the sum of intraband and interband components: $\sigma_{\rm g} = \sigma_{\textnormal{intra}} +\sigma_{\textnormal{inter}}$. More precisely, their expressions on the imaginary frequency axis are given by [\onlinecite{Chahine2017prl},\onlinecite{Falkovsky_Graphene_1},\onlinecite{Falkovsky_Graphene_2},\onlinecite{Awan_Graphene_3}]

\begin{equation}
\begin{aligned} 
\sigma_{\textnormal{intra}} (i\xi_m) &= \dfrac{8\sigma_0 k_B T}{\pi(\hbar \xi_m+\hbar/\tau)}\ln\left[ 2 \cosh\left(\dfrac{\mu}{2k_BT}\right) \right],\\ 
\sigma_{\textnormal{inter}}  (i\xi_m)&= \dfrac{\sigma_0 4\hbar \xi_m}{\pi}\int_0^{+\infty} \dfrac{G(x)}{(\hbar \xi_m)^2 + 4x^2} dx,
\label{sig_tot}
\end{aligned}
\end{equation} 
where, $\sigma_0={e^2}/({4\hbar})$, $e$  is the electron charge, $G(x) = \sinh (x/k_B T) / [\cosh (\mu/k_B T) + \cosh(x/k_B T)]$, $\tau$ the relaxation time (we use $\tau = 10^{-13}${s}).
\section{\label{reflexion_operators}Reflection operators}
In this section, we present the derivation of the reflection operators for the interacting bodies. We begin with body 1, which comprises a dielectric grating covered with graphene, as depicted in Fig. \ref{Physical_system}. The reflection operator can be directly obtained, as sub-blocs, from the scattering matrix, which has dimensions of $4(2N + 1) \times 4(2N + 1)$ and can represented as follows

\begin{equation}
\mathbb{S}=\begin{pmatrix}
\mathcal{R}^{-} & \mathcal{T}^{-}\\
\mathcal{T}^{+} & \mathcal{R}^{+}
\end{pmatrix}.
\label{R_matrix_conical}
\end{equation} 

This matrix is obtained through the use of the  FMM (see detailed calculations in the appendix) and takes the form:
\begin{equation}
\mathbb{S}=\left(\begin{array}{cc}
\mathbb{1}& \mathbb{0}  \\
\mathbb{0} &\Phi 
\end{array}\right)\mathbb{\tilde{S}}\left(\begin{array}{cc}
\mathbb{1}& \mathbb{0}  \\
\mathbb{0} &\Phi
\end{array}\right),
\label{smatrix}
\end{equation}
where
\begin{equation}
\begin{aligned}
\Phi\equiv&\left(\begin{array}{cc}
\operatorname{diag}\left(e^{i k_{zn}^{(4)} (h_1+h_2)}\right)_{n} & \mathbb{0} \\
\mathbb{0} & \operatorname{diag}\left(e^{i k_{zn}^{(4)} (h_1+h_2)}\right)_{n}
\end{array}\right),
\end{aligned}
\end{equation}
$k_{zn}^{(i)}=\sqrt{\varepsilon_{i}(\omega)\dfrac{\omega^2}{c^2}-{\bf{k}}_n^2}$ for medium $i$ (see details in the appendix), and
 \begin{equation}
\mathbb{\tilde{S}}=\mathbb{\tilde{S}}_{1} \circledast \mathbb{\tilde{S}}_{2} \circledast \mathbb{\tilde{S}}_{3} .
\end{equation}
Here, the star product operation $\mathbbm{A}= \mathbbm{B} \circledast \mathbbm{C}$ is defined as \cite{Messina2017prb}
\begin{equation}
\begin{aligned}
&\mathbbm{A}_{11} = \mathbbm{B}_{11} + \mathbbm{B}_{12}(\mathbbm{1}-\mathbbm{C}_{11}\mathbbm{B}_{22})^{-1} \mathbbm{C}_{11} \mathbbm{B}_{21},\\
&\mathbbm{A}_{12} = \mathbbm{B}_{12}(\mathbbm{1}-\mathbbm{C}_{11}\mathbbm{B}_{22})^{-1} \mathbbm{C}_{12},\\
&\mathbbm{A}_{21} = \mathbbm{C}_{21}(\mathbbm{1}-\mathbbm{B}_{22}\mathbbm{C}_{11})^{-1} \mathbbm{B}_{21},\\
&\mathbbm{A}_{22} = \mathbbm{C}_{22} + \mathbbm{C}_{21}(\mathbbm{1}-\mathbbm{B}_{22}\mathbbm{C}_{11})^{-1} \mathbbm{B}_{22}\mathbbm{C}_{12},
\label{star_prod}
\end{aligned}
\end{equation}
and
\begin{equation}
\begin{aligned}
\mathbb{\tilde{S}}_{1}&=\left(\begin{array}{ll}
\;\;\;\; \mathbb{K}_{1}^{\prime} & -\mathbb{P} \\
\mathbb{L}_{1}^{\prime} +\mathbb{M}_{1}^{\prime}  & -\mathbb{P}^{\prime}
\end{array}\right)^{-1}\left(\begin{array}{cc}
\mathbb{K}_{1} & \mathbb{P} \\
\mathbb{L}_{1} +\mathbb{M}_{1} & -\mathbb{P}^{\prime}
\end{array}\right), \;\;\;\;\;\;\;\;\;\;\;\; \\
\mathbb{\tilde{S}}_{2}&=\left(\begin{array}{cc}
\Phi_{1}& \mathbb{0} \\
\mathbb{0} & \mathbb{1}
\end{array}\right)\left(\begin{array}{ll}
-\mathbb{P} & -\mathbb{K}_{3} \\
\mathbb{P}^{\prime} & -\mathbb{L}_{3}
\end{array}\right)^{-1}\left(\begin{array}{ll}
\mathbb{P} & -\mathbb{K}_{3}^{\prime} \\
\mathbb{P}^{\prime} & -\mathbb{L}_{3}^{\prime}
\end{array}\right)\left(\begin{array}{cc}
\Phi_{1}& \mathbb{0} \\
\mathbb{0} & \mathbb{1}
\end{array}\right), \\
\mathbb{\tilde{S}}_{3}&=\left(\begin{array}{cc}
\Phi_{2} & \mathbb{0}  \\
\mathbb{0} & \mathbb{1}
\end{array}\right)\left(\begin{array}{ll}
\mathbb{K}_{3}^{\prime} & \mathbb{K}_{4} \\
\mathbb{L}_{3}^{\prime} & \mathbb{L}_{4}
\end{array}\right)^{-1}\left(\begin{array}{ll}
\mathbb{K}_{3} & \mathbb{K}_{4}^{\prime} \\
\mathbb{L}_{3} & \mathbb{L}_{4}^{\prime}
\end{array}\right)\left(\begin{array}{cc}
\Phi_{2} & \mathbb{0} \\
\mathbb{0} & \mathbb{1}
\end{array}\right).
\end{aligned}
\end{equation}
In these matrices, we have introduced
\begin{equation}
\begin{aligned}
\quad \mathbb{L}_{i}&=\sqrt{\varepsilon_{i}}\left(\begin{array}{cc}
\mathbb{B}_{xi} & \mathbb{A}_{y} \\
\mathbb{B}_{yi} & -\mathbb{A}_{x}
\end{array}\right), \quad \;\; \mathbb{L}_{i}^{\prime}=\sqrt{\varepsilon_{i}}\left(\begin{array}{cc}
\mathbb{B}_{xi} & -\mathbb{A}_{y} \\
\mathbb{B}_{yi} & \mathbb{A}_{x}
\end{array}\right), \\
 \mathbb{K}_{i}&=\left(\begin{array}{cc}
\mathbb{A}_{y} & -\mathbb{B}_{xi} \\
-\mathbb{A}_{x} & -\mathbb{B}_{yi}
\end{array}\right), 
\;\;\;\;\;\;\;\;\;\mathbb{K}_{i}^{\prime}=\left(\begin{array}{cc}
-\mathbb{A}_{y} & -\mathbb{B}_{xi} \\
\mathbb{A}_{x} & -\mathbb{B}_{yi}
\end{array}\right), \;\;\;\;\;\;\;\\
 \mathbb{M}_{i}&=\small{\sigma_{\rm g}Z_0}\left(\begin{array}{cc}
-\mathbb{A}_{x} & -\mathbb{B}_{yi} \\
-\mathbb{A}_{y} & \mathbb{B}_{xi}
\end{array}\right), 
 \;\;\mathbb{M'}_{i}=\small{\sigma_{\rm g}Z_0}\left(\begin{array}{cc}
\mathbb{A}_{x} & -\mathbb{B}_{yi} \\
\mathbb{A}_{y} & \mathbb{B}_{xi}
\end{array}\right),
\end{aligned}
\end{equation}
with 
\begin{equation}
\begin{aligned}
\;\;\;\;\;\;\;\mathbb{A}_{x}  &=\operatorname{diag}\left(\frac{k_{x n}}{k_{n}}\right)_{n}, \quad \mathbb{A}_{y}=\operatorname{diag}\left(\frac{k_{y}}{k_{n}}\right)_{n}, \\
\mathbb{B}_{xi}  &=\frac{c}{\sqrt{\varepsilon_{i}} \omega} \operatorname{diag}\left(\frac{k_{x n}}{k_{n}} k_{z n}^{(i)}\right)_{n}, \\
\mathbb{B}_{y i}  &=\frac{c}{\sqrt{\varepsilon_{i}} \omega} \operatorname{diag}\left(\frac{k_{y}}{k_{n}} k_{z n}^{(i)}\right)_{n},
\end{aligned}
\end{equation}
and
\begin{equation}
\begin{aligned}
\;\;\;\;\; \Phi_1 \equiv &e^{\mathbb{D} h_1}=\left(\begin{array}{cc}
e^{\mathbb{D}^{(11)} h_1} & \mathbb{0} \\
\mathbb{0} & e^{\mathbb{D}^{(22)} h_1}
\end{array}\right),  \\
\Phi_2\equiv&\left(\begin{array}{cc}
\operatorname{diag}\left(e^{i k_{zn}^{(3)} h_2}\right)_{n} & \mathbb{0} \\
\mathbb{0} & \operatorname{diag}\left(e^{i k_{zn}^{(3)}h_2}\right)_{n}
\end{array}\right).
\end{aligned}
\end{equation}
$Z_0$ denotes the impedance of vacuum.

The reflection matrix of body 1, denoted as $\mathcal{R}^{(1)+}$, can be derived directly from the matrix $\mathcal{R}^{-}$. Specifically, due to the distinct orientations along the z-axis, $\mathcal{R}^{(1)+}$ shares identical characteristics with $\mathcal{R}^{-}$ within the two diagonal blocks. However, a sign difference is observed within the two off-diagonal blocks, detailed in \cite{Messina2017prb}, as follows

\begin{equation}
\mathcal{R}^{(1)+} = \begin{cases}
\mathcal{R}^{-}_{p,p'} & p=p'\\
-\mathcal{R}^{-}_{p,p'} & p \neq p'.
\end{cases}
\end{equation}

For body 2, positioned at a distance $d$ from the origin, a phase shift is induced in the reflection matrix, as outlined in \cite{PhysRevA.84.042102}. This reflection matrix can be formulated as

\begin{equation}
\begin{aligned}
\left \langle p,{\textbf{k}},n|\mathcal{R}^{(2)-}(\omega)|p',{\textbf{k}}',n' \right\rangle  \\
= e^{i\left(k_{zn}^{}+k_{zn'}'\right) d}&	\left \langle p,{\textbf{k}},n|\mathcal{\tilde{R}}^{(2)-}(\omega)|p',{\textbf{k}}',n' \right\rangle.
\label{R2}
\end{aligned}
\end{equation}

Here, $\mathcal{\tilde{R}}^{(2)-}$ is diagonal with dimensions $2(2N+1) \times 2(2N+1)$ and can be easily obtained using the Fresnel coefficients:

\begin{equation}
\mathcal{\tilde{R}}^{(2)-} = {\rm diag}(\rho_{n}^{\rm TE},\rho_{n}^{\rm TM})
\label{operator_Rt2m}
\end{equation}
where 
\begin{align}
\rho_{n}^{\rm TE} &= \dfrac{k_{zn}-k_{zn}^{\rm Au}}{k_{zn}+k_{zn}^{\rm Au}} \\
\rho_{n}^{\rm TM} &= \dfrac{\varepsilon_{\rm Au}(\omega) k_{zn}-k_{zn}^{\rm Au}}{\varepsilon_{\rm Au}(\omega) k_z+k_{z}^{\rm Au}}
\label{r_Fresnel}
\end{align}
In these expressions, $k_{zn}=\sqrt{\dfrac{\omega^2}{c^2}-{\bf{k}}_n^2}$.

Finally, it is worth noting that this calculation is also valid for imaginary Matsubara frequencies by replacing $\omega$ with $i\xi_m$.
\section{\label{results}Results and discussion}
\subsection{The effect of graphene coating}

\begin{figure} [htbp]
\vspace{1em}
\centerline {\includegraphics[width=0.5\textwidth]{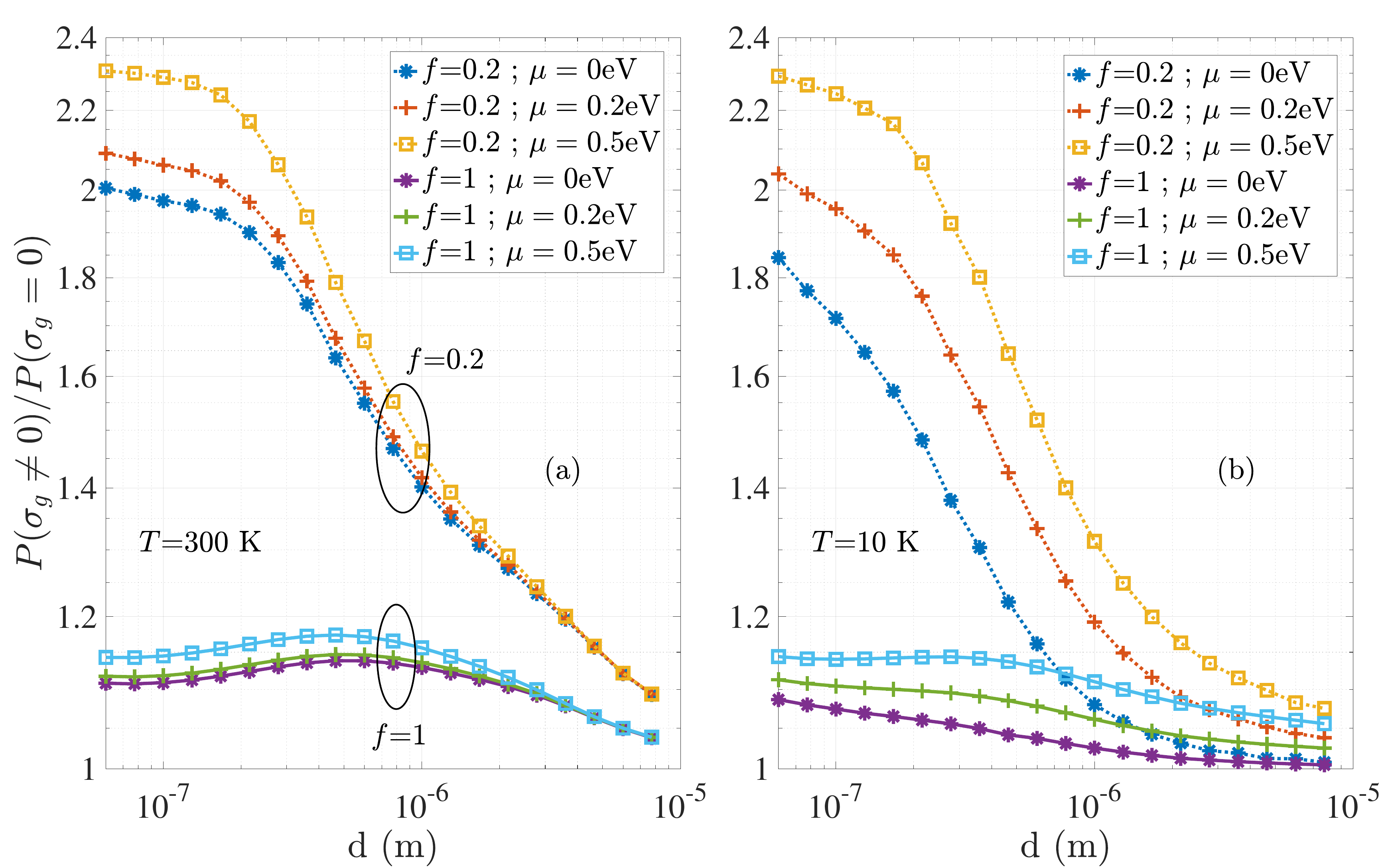}}
\caption{Ratio of the CLP with graphene to CLP without graphene for different values of $\mu$ and $f$ at $T$ = 300 K (a) and $T$ = 10 K (b). Here, $D = 1$ $\mu$m, $h_1 = 300$ nm, and $h_2 = 100$ nm.}
\label{ration_with_without}
\end{figure}

Now that we have all the necessary ingredients to calculate the Casimir-Lifshitz pressure for the configuration described previously (see section \ref{sec:Phys_sys}), let us discuss some numerical results. To begin with, we explore the impact on the CLP of covering the dielectric grating with graphene. The ratio of the pressure with graphene to that without graphene across various values of $\mu$ and $f$, for separation distances ranging from 60 nm to 8 $\mu$m, at temperatures $T = 300$ K and $T = 10$ K is depicted in Fig. \ref{ration_with_without}. 

At $T = 300$ K and for the slab configuration ($f = 1$), we observe in Fig. \ref{ration_with_without}(a), that at short distances, below 60 nm, the CLP with graphene is approximately 10\% higher than without graphene for $\mu = 0$ and $\mu = 0.2$ eV (solid lines with * and + markers), and 17\% higher for $\mu = 0.5$ eV (solid line with square marker). This enhancement peaks at $d = 500$ nm before declining as $d$ increases. Beyond this separation, the CLP enhancement is almost identical for the three values of $\mu$.
Figure \ref{ration_with_without}(b) shows the ratio when the temperature is reduced to 10 K for the same range of $d$. The enhancement remains $<$ 20\%. For $\mu$ = 0.2 eV and 0 eV, no peaks are observed. The ratio decreases to 5\% at $d \approx$ 1$\mu$m.

As the system cools down to $T = 10$ K, we observe a behavior similar to that at $T = 300$ K at short separation distances with minor variations. However, as the separation distance increases, the influence of graphene diminishes and becomes negligible, with this effect being less than 5\% for $\mu = 0$ eV and $\mu = 0.2$ eV.

On the other hand, the effect of graphene becomes more pronounced for the grating configuration. Considering $T = 300$ K and $f$=0.2, as shown Fig. \ref{ration_with_without}(a), we clearly see significant changes on the CLP, with approximately a 100\% increase for $\mu = 0$ eV (dotted line with *), 110\% for $\mu = 0.2$ eV (dotted line with +), and 130\% for $\mu = 0.5$ eV (dotted line with square) at short separation distances ({\em i.e.} below $d = 60$ nm). Similar trends persist at $T = 10$ K (Fig. \ref{ration_with_without}(b)), where the relative change remains considerable, around 85\% for $\mu = 0$ eV, 105\% for $\mu = 0.2$ eV, and 130\% for $\mu = 0.5$ eV. This enhancement decreases with increasing separation distance, reaching approximately 40\% at 1 $\mu$m for $T = 300$ K, irrespective of the chemical potential. However, at $T = 10$ K,  the impact of graphene at this separation distance varies with the chemical potential, with enhancements of 13\%, 20\%, and 32\% observed for $\mu = 0$ eV, $\mu = 0.2$ eV, and $\mu = 0.5$ eV, respectively. Additionally, the enhancement of the CLP by graphene continues to diminish with increasing separation distance at both temperatures.

\begin{figure} [htbp]
\vspace{1em}
\centerline {\includegraphics[width=0.49\textwidth]{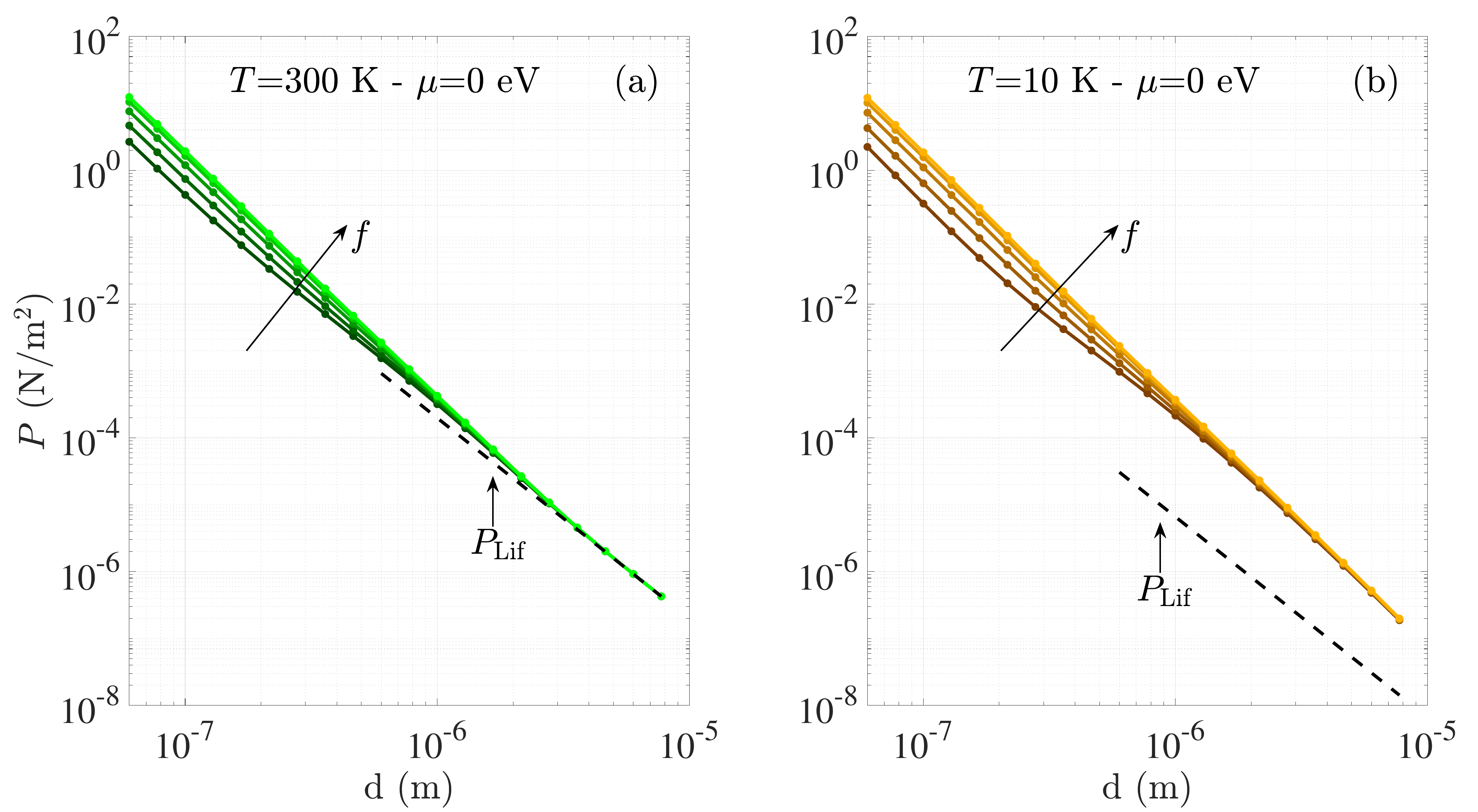}}
\caption{Casimir-Lifshitz pressure at $T = 300$ K (a) and $T = 10$ K (b), for $\mu = 0$ eV and different filling fractions $f$ = 0, 0.2, 0.5, 0.8, and 1. The Lifshitz limit for metals, $P_{\text{Lif}}$, is shown as a black dashed line.}
\label{CLP_300_10}
\end{figure}

To gain more insight into the order of magnitude of the CLP and how it varies with the geometric parameters and temperature, we plot the pressure $P$ as a function of $d$ on a log-log scale for $\mu=0$ eV and at $T=300$ K   [Fig. \ref{CLP_300_10}(a)]  and at $T=10$ K [Fig. \ref{CLP_300_10}(b)] . These plots cover distances from 60 nm to 8 $\mu$m for different filling fractions $f=0, 0.2, 0.5, 0.8$, and 1. Notably, when $f=0$, the configuration corresponds to the graphene sheet located on top of a vacuum gap 300 nm from the top surface of the SiO$_2$ layer with thickness $h_2$ = 100 nm, supported by a Silicon half-space. We observe that the CLP increases with the filling fraction $f$ regardless of temperature. Moreover, at $T=300$ K, the CLP reaches the Lifshitz limit for metals ($P_{\text{Lif}} = -\frac{k_BT\zeta(3)}{8\pi d^3}$) at nearly 6 $\mu$m (for all filling fractions), as illustrated clearly in Fig. \ref{CLP_300_10}(a). However, at $T=10$ K, depicted in Fig. \ref{CLP_300_10}(b), this limit remains unattained within the explored separation distances due to the larger thermal wavelength, $\lambda_T = \frac{\hbar c}{k_BT}$, which is approximately 229 $\mu$m. 
\subsection{Chemical potential and geometry effects}
Next, we investigate the impact of the chemical potential combined with the geometry of the grating. For that, we analyze the ratio of the CLP at $\mu = 0.2$ eV and $0.5$ eV to that at $\mu=0$ eV, for various filling fractions $f=0, 0.2, 0.5, 0.8$, and 1, and spanning distances from 60 nm to 8 $\mu$m.

\begin{figure} [htbp]
\vspace{1em}
\centerline {\includegraphics[width=0.5\textwidth]{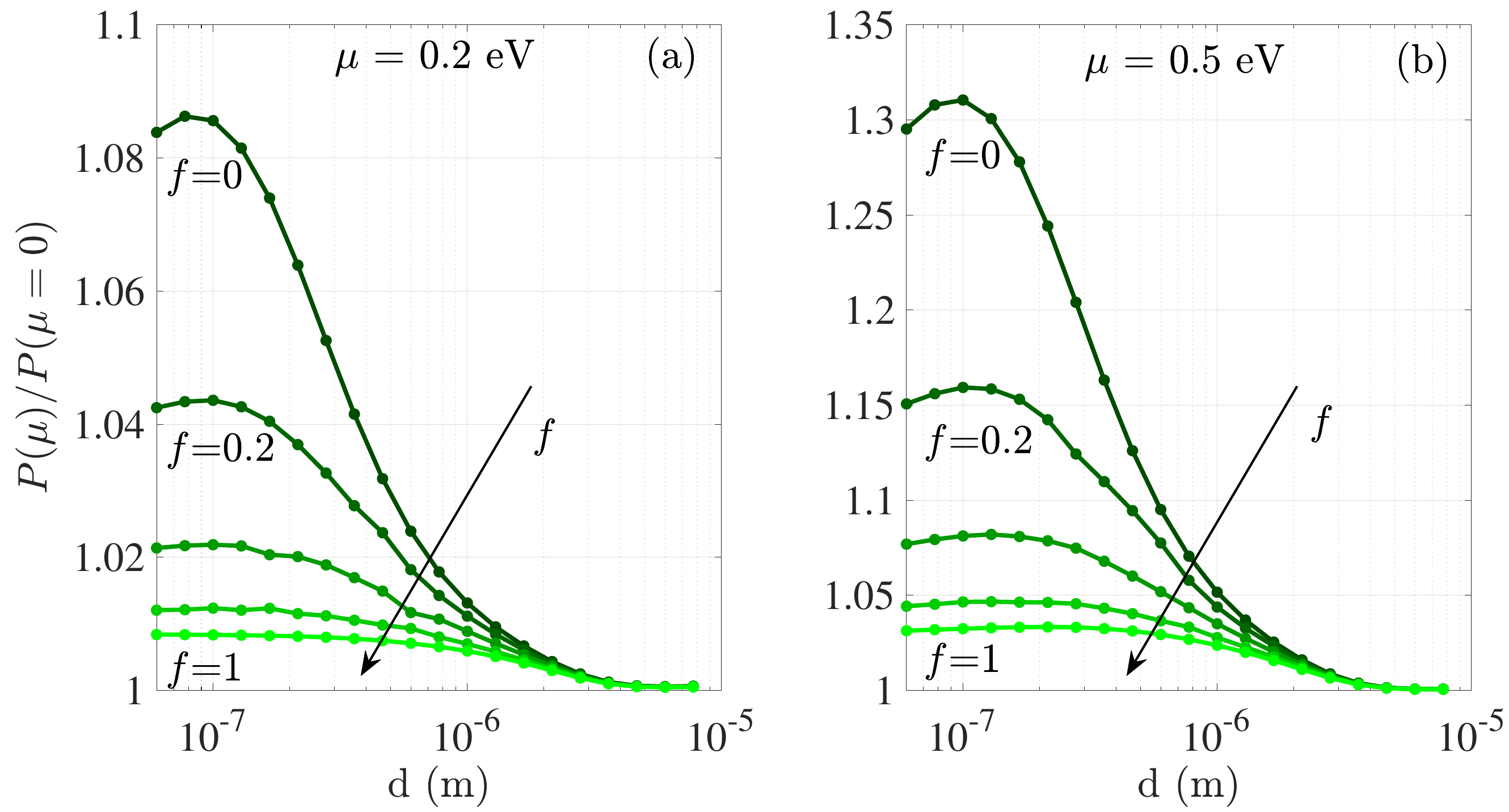}}
\caption{The ratio of CLP (calculated using Eq. \eqref{CLP}) at $\mu = 0.2$ eV (a) and $\mu = 0.5$ eV (b) to that at $\mu = 0$ eV, at $T = 300$ K, for various filling fractions $f$ = 0, 0.2, 0.5, 0.8, and 1.}
\label{Chemical_pot_T300K}
\end{figure}

At $T=300$ K and $\mu=0.2$ eV, Fig. \ref{Chemical_pot_T300K}(a) shows that having a non-zero chemical potential always increases the pressure for any value of $f$. This augmentation is larger for smaller filling fraction. The ratio peaks at around 80 nm and decreases with distance. This effect is most pronounced when $f=0$, reaching a maximum increase of $\approx$9\% and for $f=1$ (representing a slab configuration) the maximum increase is reduced to only 1\%.

As the chemical potential increases to $\mu=0.5$ eV, as shown in Fig. \ref{Chemical_pot_T300K}(b), the effect becomes more significant, with an increase of 32\% for $f=0$, 16\% for $f=0.2$ at $d \approx$ 100 nm. The maximum ratio also becomes larger for other values of $f$. 

\begin{figure} [htbp]
\vspace{1em}
\centerline {\includegraphics[width=0.5\textwidth]{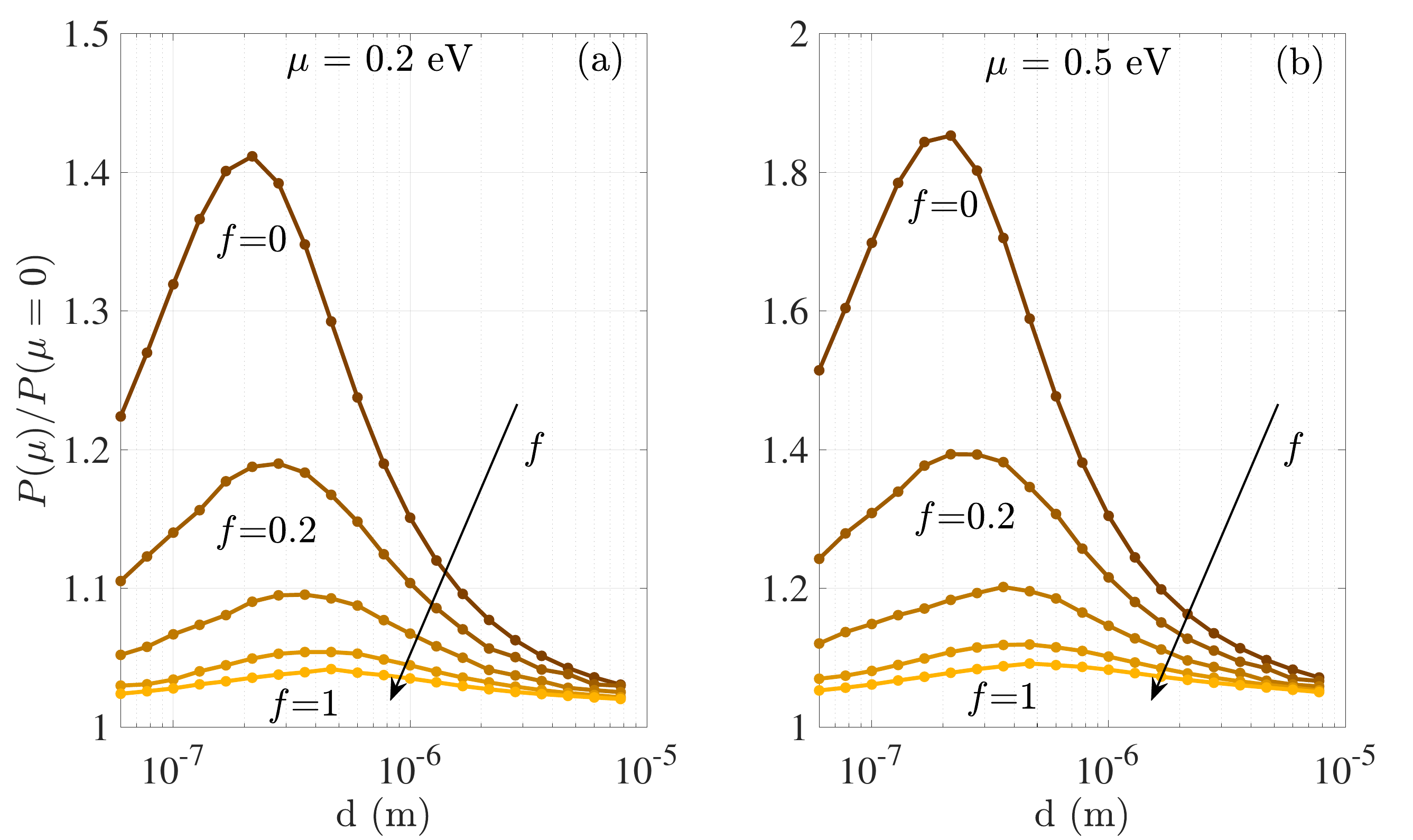}}
\caption{The ratio of CLP (calculated using Eq. \eqref{CLP}) at $\mu = 0.2$ eV (a) and $\mu = 0.5$ eV (b) to that at $\mu = 0$ eV, at $T = 10$ K, for various filling fractions $f$ = 0, 0.2, 0.5, 0.8, and 1.}
\label{Chemical_pot_T10K}
\end{figure}

Figure \ref{Chemical_pot_T10K} shows the results when the temperature is reduced to 10 K. The most notable change is that the effect of mu becomes stronger. The ratio $P(\mu)/P(\mu=0)$ becomes larger for the chosen parameters. In particular, the maximum ratio increases for all values of f compared to 300 K. For example, the peak ratio for $f$ = 0 is 1.42 and 1.85 for $\mu$ = 0.2 eV [Fig. \ref{Chemical_pot_T10K}(a)] and 0.5 eV [Fig. \ref{Chemical_pot_T10K}(b)] respectively at 10 K. Furthermore, it is apparent that the distance at which the ratio attains peak value is shifted towards larger values compared to 300K. 


\begin{figure} [htbp]
\vspace{1em}
\centerline {\includegraphics[width=0.53\textwidth]{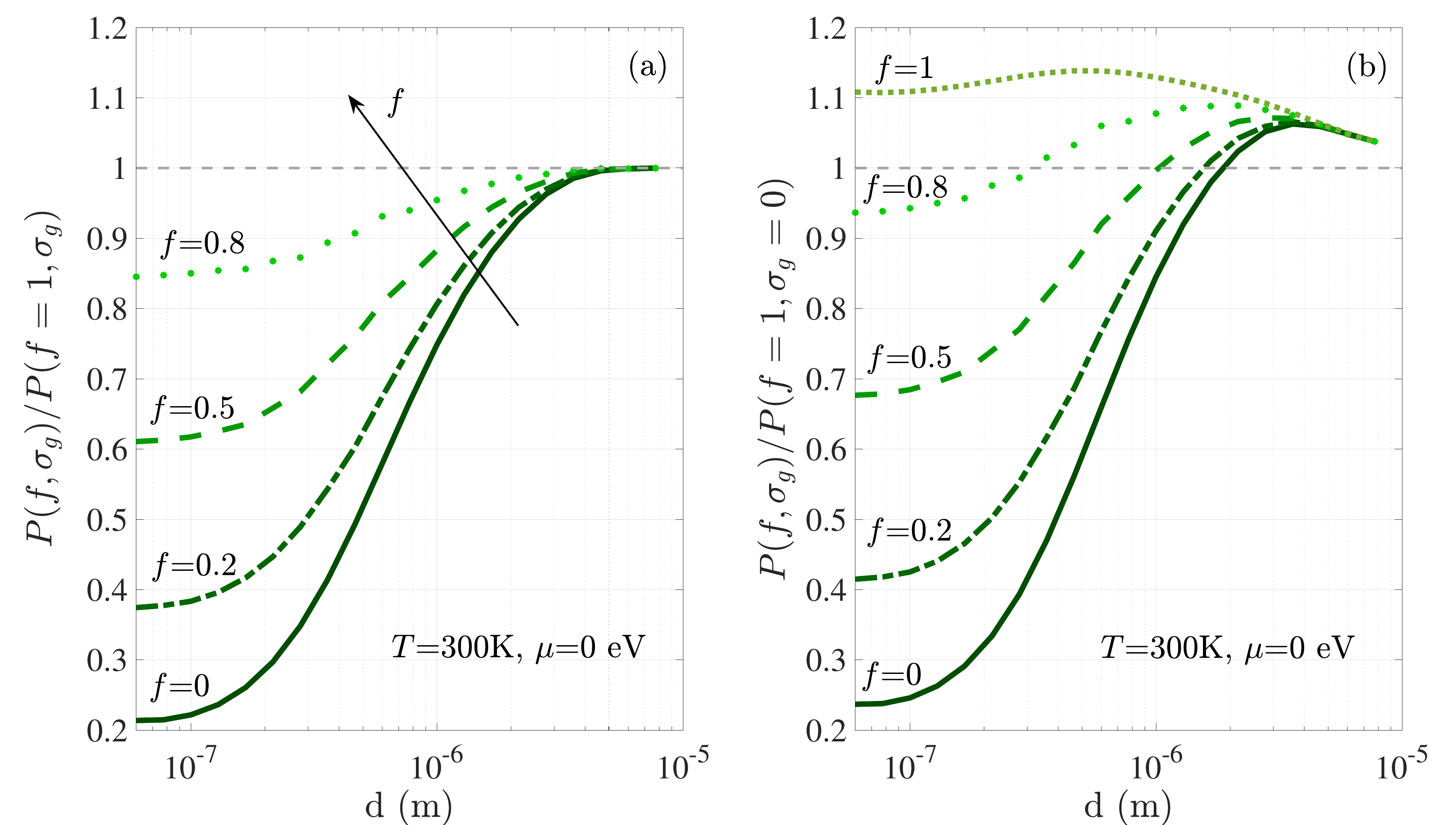}}
\caption{Normalized CLP for different filling fractions $f$ = 0, 0.2, 0.5, 0.8, and 1 (if $\sigma_g \neq 0$) relative to the CLP for $f = 1$ with graphene (a) and without graphene (b) at $T = 300$ K and for $\mu = 0$ eV.}
\label{Geom_effect}
\end{figure}

\begin{figure} [htbp]
\vspace{1em}
\centerline {\includegraphics[width=0.53\textwidth]{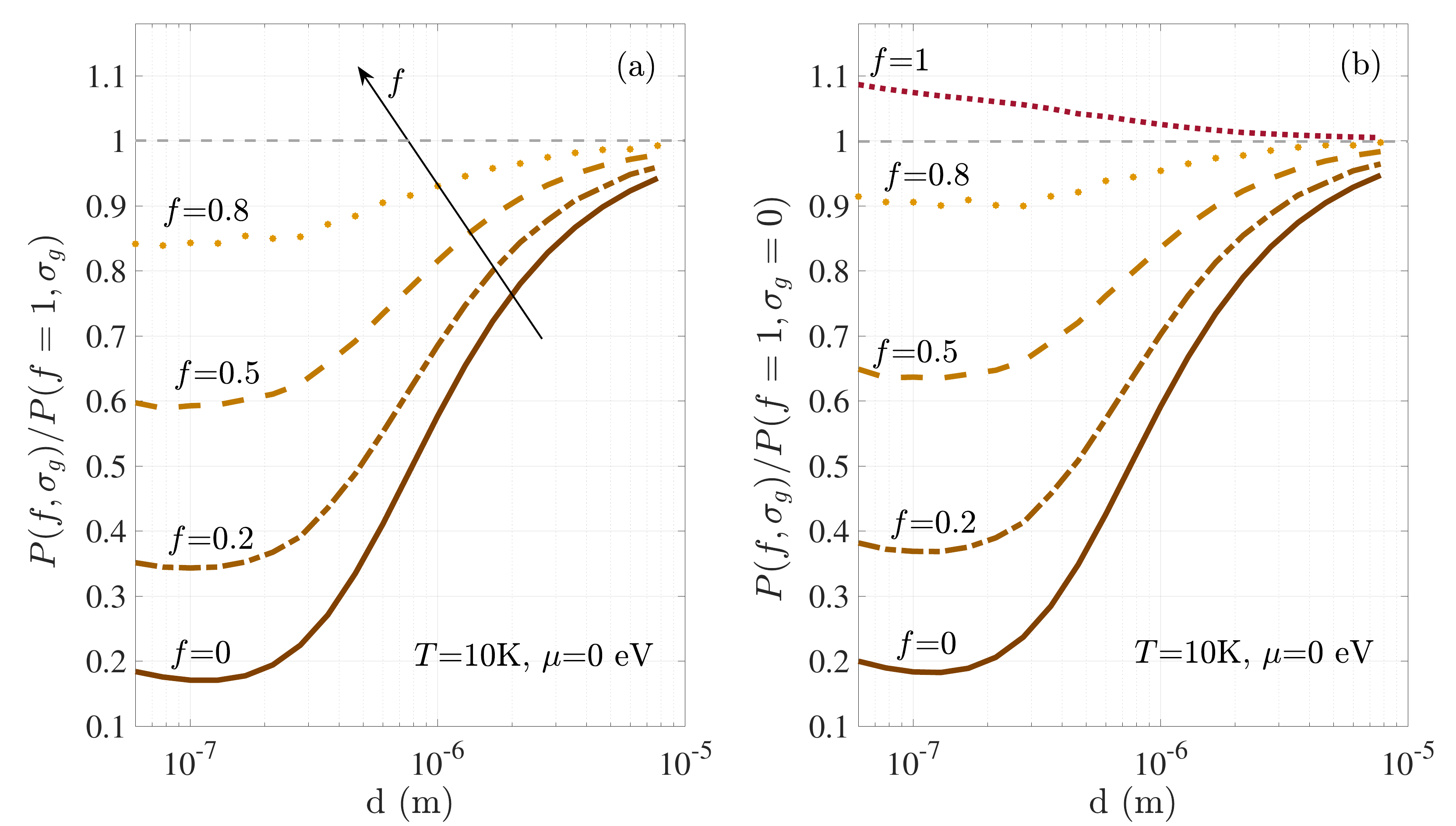}}
\caption{Normalized CLP for different filling fractions $f$ = 0, 0.2, 0.5, 0.8, and 1 (if $\sigma_g \neq 0$) relative to the CLP for $f = 1$ with graphene (a) and without graphene (b) at $T = 10$ K and for $\mu = 0$ eV.}
\label{Geom_effect_T10}
\end{figure}

To further study the impact of geometry, we analyze the ratio of the CLP for a graphene-covered dielectric grating to that of both a graphene-covered slab and a bare slab (without graphene). In Fig. \ref{Geom_effect}(a), where $T = 300$ K and $\mu = 0$ eV, the CLP between the graphene-covered grating and the gold half-space remains lower than that between a graphene-covered slab and a gold half-space when the separation distance is less than 4 $\mu$m.  Beyond 4 $\mu$m, the effect of $f$ diminishes, with the ratio approaching 1, since the two structures have the same asymthotique Lifshitz limit. Similarly, at $T$ = 10 K, as depicted in Fig. \ref{Geom_effect_T10}(a), we observe a comparable trend with a greater saturation distance.

On the other hand, the CLP for the graphene-coated dielectric grating ($f \neq$ 1) is initially lower than that for the bare slab at small separation distances, as illustrated in Fig. \ref{Geom_effect}(b). However, as the separation distance $d$ increases, this trend reverses. For $f = 0.8$, the CLP exceeds that of the bare slab beyond 300 nm, while for $f = 0.5$, this occurs beyond 1 $\mu$m. As $f$ increases, the cross-over distance shifts to larger values. The crossover no longer occurs for $T = 10$ K, as it is clear from Fig. \ref{Geom_effect_T10}(b) where the CLP for the graphene-coated grating ($f \neq$ 1) is lower that that of the bare slab. It is noteworthy that when the slab is covered with graphene ($f=1$), the pressure is always higher compared to that without graphene whether at $T = 300$ K or $T = 10$ K.
\subsection{Thermal effect and Casimir force gradient}
\begin{figure} [htbp]
\vspace{1em}
\centerline {\includegraphics[width=0.54\textwidth]{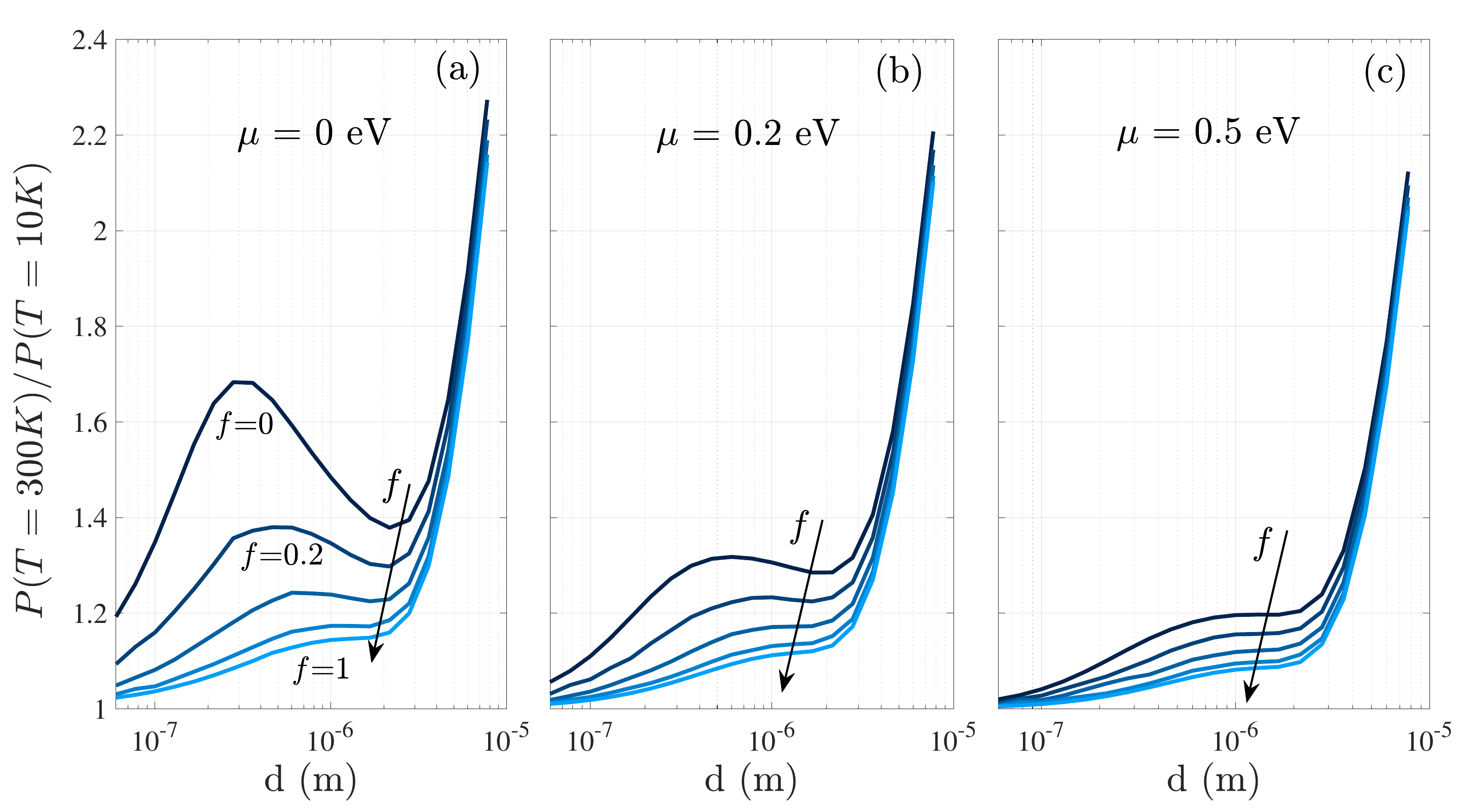}}
\caption{Ratio of the CLP at $T = 300$ K to that at $T = 10$ K for $\mu = 0$ eV (a), $\mu = 0.2$ eV (b), and $\mu = 0.5$ eV (c), for different filling fractions $f$ = 0, 0.2, 0.5, 0.8, and 1.}
\label{Thermal_effect}
\end{figure}

Now we explore the thermal impact of graphene by examining the ratio of the CLP at $T = 300$ K to that at $T = 10$ K. Initially, for $\mu = 0$ eV and $f = 0$, a notable thermal effect emerges around 300 nm, where the ratio of approximately 1.7 (Fig. \ref{Thermal_effect}(a)). As we increase the filling fraction, this peak gradually diminishes in height and shifts towards larger separation distances. For instance, the local peak decreases to approximately 1.4 for $f = 0.2$ (at $d$=500 nm) and 1.15 at $d=1\mu$m for $f = 1$. After this peak (for every $f$), the ratio decreases, before eventually experiencing a sharp increase for distances $\approx 6 \mu$m. The sharp rise is expected because the pressure at $T = 300$ K reaches the Lifshitz limit (see Fig. \ref{CLP_300_10}(a)) while remaining below it at $T = 10$ K (see Fig. \ref{CLP_300_10}(b)). In this limiting case, using the expression $P_{\text{Lif}} = k_BT\zeta(3)/(8\pi d^3)$ yields a ratio of 30. Consequently, it is expected that the ratio continues to increase until reaching the thermal wavelength, approximately 229 $\mu$m, beyond which the value levels off. 

Moreover, this thermal effect decreases with the chemical potential, as demonstrated in Fig. \ref{Thermal_effect}(b) and \ref{Thermal_effect}(c) for $\mu = 0.2$ and 0.5, respectively. This is due to the fact that the graphene conductivity \eqref{sig_tot} only depends weakly on the temperature $T\approx 10^{-3} eV$  when $\mu$ is larger than $T$.

Finally, we consider the gradient of the Casimir force between a gold sphere and a dielectric grating covered with graphene. According to the Proximity Force Approximation (PFA) \cite{PFA_gradF,Bimonte2018prd}, the Casimir pressure $P$ between two parallel plates is proportional to the gradient of the Casimir force in the sphere-plate geometry, and it can be expressed as $F' = 2\pi R P$, where $R$ is the sphere radius. The latter configuration is used in experiments to avoid the difficulty in keeping two planar bodies parallel at short separations. We considered a gold sphere with a radius of 40 $\mu$m at $T = 300$ K and $T = 10$ K, and calculated the gradient over distances ranging from 60 nm to 1 $\mu$ m for two configurations: a slab ($f = 1$) and a grating with $f = 0.2$ and $\mu = 0$ eV.

Our analysis revealed that the effect of the graphene on the force gradient is significantly more pronounced for a grating compared to solid SiO$_2$, particularly at short separation distances for both $T$ = 10 K and 300 K. the force gradient at $d $= 200 nm increases by 94\% and 55\% respectively for $T$ = 300 K and 10K when graphene is added to an SiO$_2$ grating with $f$ = 0.2.


\begin{figure} [htbp]
\vspace{1em}
\centerline {\includegraphics[width=0.5\textwidth]{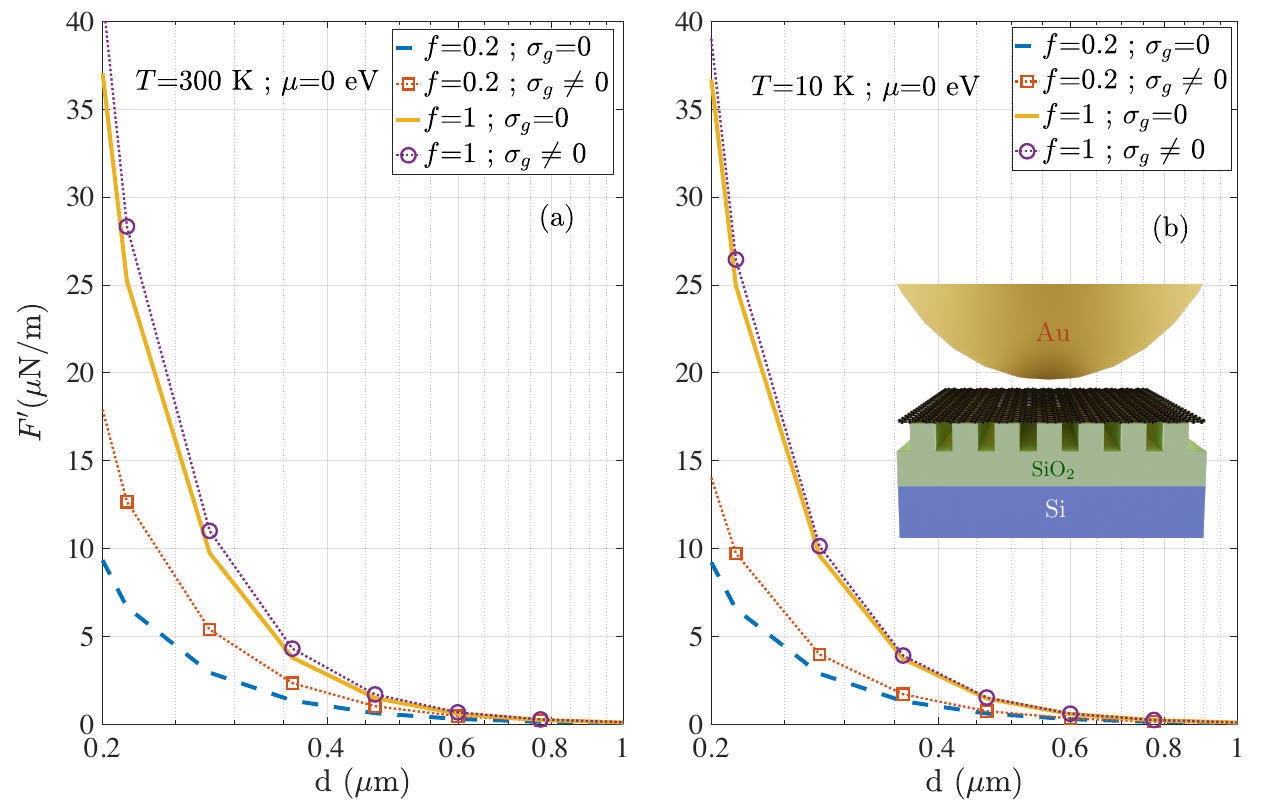}}
\caption{Gradient of the Casimir force between a gold sphere with radius $R$ = 40 $\mu$m and a graphene-coated fused silica grating with the following parameters: period $D$ = 1 $\mu$m, depth $h_1$ = 300 nm, and slab thickness $h_2$ = 100 nm, at $T = 300$ K (a) and $T = 10$ K (b). for different values of $\mu$ and $f$.}
\label{gradient_force}
\end{figure}

\section{Conclusion}
In conclusion, we have investigated the Casimir-Lifshitz pressure between a graphene-coated dielectric grating and a gold half-space using an adapted FMM. Our results reveal that covering the dielectric grating with graphene leads to a significant increase in pressure, with enhancements up to 130\% for $\mu = 0.5$ eV and $f=0.2$, compared to only 17\% when covering a slab with graphene, over a wide range of temperatures from 10K to 300K. Additionally, we have shown that the CLP depends strongly on the chemical potential of graphene, with maximal effects observed at lower filling fractions ($f$ = 0.2 in our case) for both $T $= 300 K and 10 K. Notably, our findings indicate that the pressure for a graphene-coated dielectric grating can surpass that of a bare slab at $T = 300$ K at certain separation distances even when $f$ is small. Furthermore, we have identified a thermal effect between 300 nm and 400 nm, which diminishes with increasing filling factor and chemical potential. Lastly, we have presented the Casimir force gradient between a gold sphere and a graphene-coated dielectric grating, a configuration that is used for experimental measurements.
\begin{acknowledgments}

The work described in this paper was supported by a grant "CAT" from the ANR/RGC Joint Research Scheme sponsored by the French National Research Agency (ANR) and the Research
Grants Council (RGC) of the Hong Kong Special Administrative Region{, China (Project No. A-HKUST604/20).} \\

\end{acknowledgments}

\section*{}
\section*{\label{sMethod}APPENDIX: SCATTERING MATRIX CALCULATION FOR GRAPHENE-COVERED GRATING USING THE FOURIER MODAL METHOD}
\begin{figure} [htbp]
\vspace{1em}
\centerline {\includegraphics[width=0.5\textwidth]{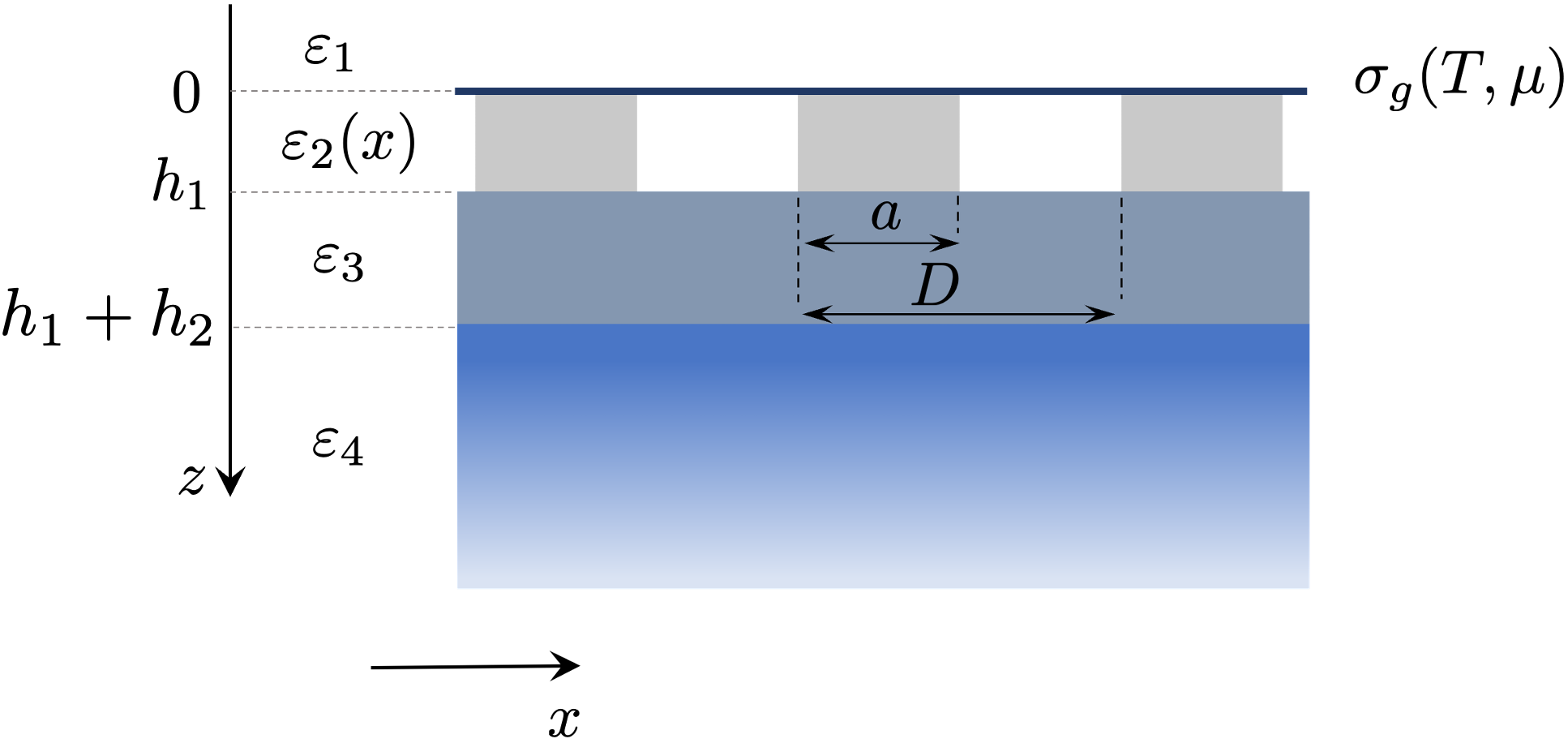}}
\caption{Geometry of the FMM calculation: A dielectric grating characterized by dielectric permittivity $\varepsilon_2(x)$, with period $D$, width $a$, and depth $h_1$, covered with a graphene sheet and placed on a top of a slab with dielectric permittivity $\varepsilon_3$ and thickness $h_2$, which in turn rests on a dielectric half-space with dielectric permittivity $\varepsilon_4$.}
\label{structure}
\end{figure}

In this appendix, we will describe the calculation of the scattering matrix using the Fourier Modal Method for dielectric gratings coated with graphene, employing a zero-thickness model. This approach simplifies the representation of graphene and facilitates its direct inclusion into the boundary conditions of the FMM.
The structure under study is depicted in Fig. \ref{structure}, where a graphene sheet is positioned at $z=0$. Four distinct zones can be distinguished : (i) zone 1 ($z < 0$) is characterized by $\varepsilon_1$, (ii) zone 2 ($0 < z < h_1$) contains  a lamellar grating with a period $D$ along the $x$-axis, a relative dielectric permittivity defined denoted $\varepsilon_2(x)$ and a thickness $h_1$, (iii) zone 3 $(h_1 < z < h_1 + h_2)$, is a homogeneous slab with permittivity $\varepsilon_3$ and (iv) zone 4 ($z > h_1 + h_2$) is a semi-infinite medium characterized by $\varepsilon_4$.

\subsection{Electromagnetic Fields}
Due to periodicity along the $x-$axis, the electromagnetic fields in each homogeneous zone ($i =1, 3, 4$) can be expanded into the so-called Rayleigh expansion as follows:

\begin{equation}
\left\{
\begin{aligned}
{\bf E}^{(i)}({\bf R}, \omega) &= \sum_{p, \phi} \int_{-\frac{\pi}{D}}^{\frac{\pi}{D}} \dfrac{\mathrm{d} k_x}{2 \pi} \sum_{n \in \mathbb{Z}} \int_{-\infty}^{+\infty} \dfrac{\mathrm{d} k_y}{2 \pi}   \\
&\times e^{i {\bf K}_n^{(i) \phi}\cdot \bf{R}} 
\hat{{\bf e}}_{p}^{(i)\phi}({\bf k}_n,\omega) E_p^{(i) \phi}({\bf k}_n,\omega),\\
{\bf B}^{(i)}({\bf R}, \omega) &= \frac{\sqrt{\varepsilon_i(\omega)}}{c} \sum_{p, \phi} \int_{-\frac{\pi}{D}}^{\frac{\pi}{D}} \dfrac{\mathrm{d} k_x}{2 \pi} \sum_{n \in \mathbb{Z}} \int_{-\infty}^{+\infty} \dfrac{\mathrm{d} k_y}{2 \pi}   \\
&\times e^{i {\bf K}_n^{(i) \phi} \cdot \bf{R}} (-1)^p {\hat{{\bf e}}_{S(p)}^{(i) {\bf \phi}}}({\bf k}_n,\omega) E_p^{(i) \phi}({\bf k}_n,\omega),
\end{aligned}
\right.
\label{E_B_homog}
\end{equation}
where $p$ denotes the index of polarization ($p$=1,2 for TE and TM respectively), ${\bf R} = ({\bf r},z)$, $S(p)$ is a function ($S(1)=2$ and $S(2)=1$), and  

\begin{equation}
\left\{
\begin{aligned}
& \hat{\bf{e}}_{\rm TE}^{(i) \phi}({\bf{k}}_{n},\omega)=\frac{1}{k_n}(-k_y \hat{\bf{e}}_{x}+k_{xn}\hat{\bf{e}}_{y}),\\
& \hat{\bf{e}}_{\rm TM}^{(i) \phi}({\bf{k}}_{n},\omega)=\frac{c}{\omega \sqrt{\varepsilon_i(\omega)}}(-k_n  \hat{\bf{e}}_{z} + \phi k_{zn}^i \hat{\bf{k}}_{n}).
\label{TE_TM_basis_vectors}
\end{aligned}
\right.
\end{equation}
Here, $\hat{\bf{e}}_{x}$, $\hat{\bf{e}}_{y}$ and $\hat{\bf{e}}_{z}$ are unit vectors in the $(x, y, z)$ Cartesian basis, $\hat{\bf{k}}_{n}={\bf{k}}_{n}/k_n$, $\phi$ {is} the direction of propagation of the waves $(+, -)$ along the $z$-axis for the incident and the reflected fields, respectively, and ${\bf K}_n^{(i) \phi}= ({\bf k}_n,\phi k_{zn}^{(i)})$.

For the different amplitudes in these zones, we use the notations :
\begin{equation}
\left\{
\begin{aligned}
E^{(1)+} &= I, \; \; \; \; E^{(1)-} = R,\\
E^{(3)+} &= C, \; \; \; \; E^{(3)-} = C',\\
E^{(4)+} &= T, \; \; \; \;E^{(4)-} = J,\\
\end{aligned}
\right.
\end{equation}

In the periodic region (zone 2), the electric field assumes the following form:

\begin{equation}
{\bf E}^{(2)}({\bf R}, \omega) =  \int_{-\frac{\pi}{D}}^{\frac{\pi}{D}} \dfrac{\mathrm{d} k_x}{2 \pi} \sum_{n \in \mathbb{Z}} \int_{-\infty}^{+\infty} \dfrac{\mathrm{d} k_y}{2 \pi} e^{i {\bf k}_n \cdot \bf{r}}  {\bf E}^{(2)} (z,{\bf k}_n,\omega).
\label{E_Grating}
\end{equation}
In the actual computation, we will truncate the sum ranging from $-\infty$ to $\infty$, retaining only $2N+1$ Fourier harmonics, $N$ is termed the truncation order. In order to obtain the amplitudes ${\bf E}^{(2)} (z,{\bf k}_n,\omega)$, we have to solve Maxwell's equations within this zone:

\begin{equation}
\left\{
\begin{aligned}
\partial_y E_z- \partial_z E_y = i  k_0 \tilde{H}_x\\
\partial_z E_x- \partial_x E_z = i k_0 \tilde{H}_y\\
\partial_x E_y- \partial_y E_x = i k_0 \tilde{H}_z\\
\end{aligned}
\right.
\left\{
\begin{aligned}
\partial_y \tilde{H}_z- \partial_z \tilde{H}_y = - i \varepsilon k_0 E_x\\
\partial_z \tilde{H}_x- \partial_x \tilde{H}_z = - i \varepsilon k_0 E_y\\
\partial_x \tilde{H}_y- \partial_y \tilde{H}_x = - i \varepsilon k_0 E_z,\\
\end{aligned}
\right.
\label{Maxwell_fmm}
\end{equation}
with $k_0 = \omega/c$, and $\tilde{H}=Z_0 H$.

Eliminating the $z$ components from Eq. \eqref{Maxwell_fmm}, we obtain two systems of equations involving only the parallel components over which the boundary conditions hold:

\begin{equation}
\begin{aligned}
\partial_z \begin{pmatrix}
E_{x}  \\
E_{y}
\end{pmatrix}
=\begin{pmatrix}
-\dfrac{i}{k_0}\partial_x \dfrac{1}{\varepsilon(x)}\partial_y &i k_0 + \dfrac{i}{k_0}\partial_x \dfrac{1}{\varepsilon(x)}\partial_x \\
-ik_0-\dfrac{i}{k_0}\partial_y \dfrac{1}{\varepsilon(x)}\partial_y & \dfrac{i}{k_0}\partial_y \dfrac{1}{\varepsilon(x)}\partial_x \\
\end{pmatrix}
\\ \times \begin{pmatrix}
 \tilde{H}_x  \\
 \tilde{H}_y
\end{pmatrix}
\end{aligned}
\end{equation}

\begin{equation}
\begin{aligned}
\partial_z \begin{pmatrix}
 \tilde{H}_x  \\
 \tilde{H}_y
\end{pmatrix}
=\begin{pmatrix}
 \dfrac{i}{k_0} \partial_x\partial_y&-i k_0 \varepsilon(x)- \dfrac{i}{k_0}\partial_x \partial_x  \\
i k_0 \varepsilon(x)+ \dfrac{i}{k_0}\partial_y \partial_y &   -\dfrac{i}{k_0} \partial_x\partial_y\\
\end{pmatrix}
\\ \times \begin{pmatrix}
E_{x}  \\
E_{y}
\end{pmatrix}
\end{aligned}
\end{equation}

Subsequently, following the method elaborated in more detail, for instance, in \cite{FMM_Guizal}, we transform these equations into Fourier space and get :

\begin{equation}
\left\{
\begin{aligned}
 \partial_z \bm{ \mathcal{E}}
=\mathbb{F}\bm{ \mathcal{\tilde{H}}}\\
\partial_z \bm{ \mathcal{\tilde{H}}}=\mathbb{G} \bm{ \mathcal{E}}
\end{aligned}
\right.
\label{Coupled}
\end{equation}

where  $\bm{ \mathcal{E}} =[\bm{ \mathcal{E}}_x, \bm{ \mathcal{E}}_y]^t$ and $\bm{ \mathcal{\tilde{H}}}=[\bm{ \mathcal{\tilde{H}}}_x, \bm{ \mathcal{\tilde{H}}}_y]^t$ are vectors containing the Fourier components of the electric and magnetic parallel components of the field in the grating zone. Moreover,  

\begin{equation}
\mathbb{F} =\begin{pmatrix}
\dfrac{i\beta}{k_0}\alpha\left\llbracket \varepsilon \right\rrbracket^{-1} &i k_0  \mathbbm{1}- \dfrac{i\alpha}{k_0}\left\llbracket \varepsilon \right\rrbracket^{-1} \alpha \\
-i k_0  \mathbbm{1}+ \dfrac{i\beta^2}{k_0}\left\llbracket \varepsilon \right\rrbracket^{-1}  & -\dfrac{i\beta}{k_0}\left\llbracket \varepsilon \right\rrbracket^{-1} \alpha \\
\end{pmatrix} ,
\end{equation}

and

\begin{equation}
\mathbb{G}=\begin{pmatrix}
-\dfrac{i\beta}{k_0}\alpha&- i k_0 \left\llbracket \varepsilon \right\rrbracket + \dfrac{i\alpha^2}{k_0}\\
i k_0 \left\llbracket \varepsilon ^{-1} \right\rrbracket^{-1} - \dfrac{i\beta^2}{k_0}  & \dfrac{i\beta}{k_0}\alpha \\
\end{pmatrix}.
\end{equation}

Here, $\left\llbracket \varepsilon \right\rrbracket $ is the Toeplitz matrix such that $\left\llbracket \varepsilon \right\rrbracket_{mn}= \varepsilon_{m-n}$, $\beta$ is a scalar, and $\alpha = {\rm diag}(k_{xn})_n$. 

Using the coupled equations \eqref{Coupled}, we deduce the equation for $\bm{\mathcal{E}}$ alone :
\begin{equation}
\partial^2_z  \bm{\mathcal{E}} = \mathbb{F}\mathbb{G}\bm{\mathcal{E}} =  \mathbb{P}\mathbb{D}^2\mathbb{P}^{-1}\bm{\mathcal{E}},
\label{egn_fmm}
\end{equation}
where $\mathbb{P}$ and $\mathbb{D}^2$ contain, respectively, the eigenvectors and eigenvalues of the matrix $\mathbb{FG}$, each having dimensions of $2(2N + 1) \times 2(2N + 1)$ and take the following form:
\begin{equation}
\mathbb{P} = \begin{pmatrix}
\mathbb{P} ^{(11)} & \mathbb{P} ^{(12)}\\
\mathbb{P} ^{(21)} & \mathbb{P} ^{(22)}\\
\end{pmatrix}, 
\mathbb{D} = \begin{pmatrix}
\mathbb{D} ^{(11)} &\mathbb{0} \\
\mathbb{0}  & \mathbb{D} ^{(22)}\\
\end{pmatrix}.
\end{equation}

The solution of the set of equations \eqref{Coupled} can then be written : 
\begin{equation}
\left\{
\begin{aligned}
\bm{ \mathcal{E}}&=\mathbb{P}(e^{\mathbb{D}z}\bm{ \mathcal{A}}+e^{\mathbb{-D}z}\bm{ \mathcal{B}} )\\
\bm{ \mathcal{\tilde{H}}}&=\mathbb{P'}(e^{\mathbb{D}z}\bm{ \mathcal{A}}-e^{\mathbb{-D}z}\bm{ \mathcal{B}}),
\end{aligned}
\right.
\label{Fourier_Toep}
\end{equation}
where $\bm{\mathcal{A}} = [A_{xn}, A_{yn}]^t$ and $\bm{\mathcal{B}}= [B_{xn}, B_{yn}]^t$ ($n \in [-N,N]$) are arbitrary constant vectors, and $\mathbb{P'} = \mathbb{F}^{-1}\mathbb{P}\mathbb{D}$.
\subsection{Boundary conditions}
After expressing the fields in all zones, we are going to enforce the boundary conditions at their interfaces. For both analytical and numerical convenience, we introduce an additional phase factor in the expressions of the fields within zones 3 and 4. Specifically in zone 3, we take the $z$ dependance of the fields under the form $e^{ik^{(i)\phi} (z - h_1)}$, while in zone 4, we we take it under the form $e^{ik^{(i)\phi} (z - h_1 - h_2)}$. These adjustments facilitate the calculations and can be easily reverted at the conclusion of the analysis. For this reason, we will first obtain the scattering matrix with this simplified phase $\mathbb{\tilde{S}}$, and then multiply it by the accumulated phase to obtain the actual scattering matrix $\mathbb{S}$ (see Eq. \eqref{smatrix}).\\

Using the zero thickness model for the graphene sheet and applying the boundary conditions for the electric field components  at the interface ($z$=0) between zones 1 and 2, in the cartesian coordinates system, we obtain (after projection on the Fourier basis):

\begin{widetext}
\begin{equation}
\left(\begin{array}{c}
-\frac{k_{y}}{k_{n}}\left(I_{1n}+R_{1n}\right)+\frac{c}{\sqrt{\varepsilon_{1}} \omega} k_{zn}^{(1)} \frac{k_{xn}}{k_{n}}\left(I_{2n}-R_{2n}\right)  \\
\frac{k_{xn}}{k_{n}}\left(I_{1n}+R_{1n}\right)+\frac{c}{\sqrt{\varepsilon_{1}} \omega} k_{zn}^{(1)} \frac{k_{y}}{k_{n}}\left(I_{2n}-R_{2n}\right)
\end{array}\right)=\left(\begin{array}{l}
\mathbb{P}_{nm}^{(11)}\left(A_{xm}+B_{xm}\right)+\mathbb{P}_{n m}^{(12)}\left(A_{ym}+B_{ym}\right) \\
\mathbb{P}_{n m}^{(21)}\left(A_{xm}+B_{xm}\right)+\mathbb{P}_{n m}^{(22)}\left(A_{ym}+B_{ym}\right)
\end{array}\right).
\end{equation}

Doing the same for the magnetic field components, we have: 

\begin{equation}
\begin{aligned}
\left(\begin{array}{l}
\mathbb{P}_{n m}^{\prime(11)}\left(A_{xm}-B_{xm}\right)+\mathbb{P'}_{n m}^{(12)}\left(A_{ym}-B_{y m}\right) +\frac{c}{\omega} k_{zn}^{(1)} \frac{k_{xn}}{k_{n}}\left(I_{1n}-R_{1n}\right)+\sqrt{\varepsilon_{1}} \frac{k_{y}}{k_{n}}\left(I_{2n}+R_{2n}\right)  \\
\mathbb{P}_{n m}^{\prime(21)}\left(A_{xm}-B_{xm}\right)+\mathbb{P}_{n m}^{\prime(22)}\left(A_{ym}-B_{y m}\right) +\frac{c}{\omega} k_{zn}^{(1)} \frac{k_{y}}{k_{n}}\left(I_{1n}-R_{1n}\right)-\sqrt{\varepsilon_{1}} \frac{k_{xn}}{k_{n}}\left(I_{2n}+R_{2n}\right)
\end{array}\right)\\
=\sigma_{\rm g}Z_0\left(\begin{array}{c}
\frac{k_{xn}}{k_{n}}\left(I_{1n}+R_{1n}\right)+\frac{c}{\sqrt{\varepsilon_{1}} \omega} k_{zn}^{(1)} \frac{k_{y}}{k_{n}}\left(I_{2n}-R_{2n}\right)\\
\frac{k_{y}}{k_{n}}\left(I_{1n}+R_{1n}\right)-\frac{c}{\sqrt{\varepsilon_{1}} \omega} k_{zn}^{(1)} \frac{k_{xn}}{k_{n}}\left(I_{2n}-R_{2n}\right) 
\end{array}\right).
\end{aligned}
\end{equation}

At the interface $z=h_1$ we have the following equations for the electric field\\

\begin{equation}
\begin{aligned}
\;\;\;\;\;\;\; \;\;\;\;\;\;\;\;\;\;\;\;\;\;\;\;\;\;\;\;\;&\left(\begin{array}{c}
-\frac{k_{y}}{k_{n}}\left(C_{1n}+C_{1n}^{\prime}\right)+\frac{c}{\sqrt{\varepsilon_{3}} \omega} k_{zn}^{(3)} \frac{k_{xn}}{k_{n}}\left(C_{2n}-C_{2n}^{\prime}\right)  \\
\frac{k_{xn}}{k_{n}}\left(C_{1n}+C_{1n}^{\prime}\right)+\frac{c}{\sqrt{\varepsilon_{3}} \omega} k_{zn}^{(3)} \frac{k_{y}}{k_{n}}\left(C_{2n}-C_{2 n}^{\prime}\right)
\end{array}\right)\\
\;\;\;\;\;\;\;\;\;\;\;\;\;\;&=\left(\begin{array}{l}
\mathbb{P}_{n m}^{(11)}\left(e^{\mathbb{D}_{m m}^{(11)} h_1} A_{xm}+e^{-\mathbb{D}_{m m}^{(11)} h_1} B_{x m}\right)+\mathbb{P}_{n m}^{(12)}\left(e^{\mathbb{D}_{m m}^{(22)} h_1} A_{ym}+e^{-\mathbb{D}_{m}^{(22)} h_1} B_{y m}\right) \\
\mathbb{P}_{n m}^{(21)}\left(e^{\mathbb{D}_{m m}^{(11)} h_1} A_{x, m}+e^{-\mathbb{D}_{m m}^{(11)} h_1} B_{xm}\right)+\mathbb{P}_{n m}^{(22)}\left(e^{\mathbb{D}_{m m}^{(22)} h_1} A_{y m}+e^{-\mathbb{D}_{m m}^{(22)} h_1} B_{y m}\right)
\end{array}\right),
\end{aligned}
\end{equation}
and the boundary conditions for the magnetic field are as follows:

\begin{equation}
\begin{aligned}
& \left(\begin{array}{l}
-\frac{c}{\omega} k_{z n}^{(3)} \frac{k_{xn}}{k_{n}}\left(C_{1 n}-C_{1 n}^{\prime}\right)-\sqrt{\varepsilon_{3}} \frac{k_{y}}{k_{n}}\left(C_{2n}+C_{2 n}^{\prime}\right) \\
-\frac{c}{\omega} k_{zn}^{(3)} \frac{k_{y}}{k_{n}}\left(C_{1n}-C_{1 n}^{\prime}\right)+\sqrt{\varepsilon_{3}} \frac{k_{xn}}{k_{n}}\left(C_{2 n}+C_{2 n}^{\prime}\right)
\end{array}\right) \\
& \quad=\left(\begin{array}{l}
\mathbb{P}_{n m}^{\prime(11)}\left(e^{\mathbb{D}_{m m}^{(11)} h_1} A_{x m}-e^{-\mathbb{D}_{m m}^{(11)} h_1} B_{x m}\right)+\mathbb{P'}_{n m}^{(12)}\left(e^{\mathbb{D}_{m m}^{(22)} h_1} A_{ym}-e^{-\mathbb{D}_{m m}^{(22)} h_1} B_{y m}\right) \\
\mathbb{P}_{n m}^{\prime(21)}\left(e^{\mathbb{D}_{m m}^{(11)} h_1} A_{x m}-e^{-\mathbb{D}_{m m}^{(11)} h_1} B_{x m}\right)+\mathbb{P}_{n m}^{\prime(22)}\left(e^{\mathbb{D}_{m m}^{(22)} h_1} A_{y m}-e^{-\mathbb{D}_{m m}^{(22)} h_1} B_{y m}\right)
\end{array}\right).
\end{aligned}
\end{equation}

Likewise, at the final interface, $z=h_1+h_2$, we obtain
\begin{equation}
\begin{aligned}
\left(\begin{array}{c}
-\frac{k_{y}}{k_{n}}\left(e^{i k_{zn}^{(3)} h_2} C_{1n}+e^{-i k_{zn}^{(3)} h_2} C_{1n}^{\prime}\right)+\frac{c}{\sqrt{\varepsilon_{3}} \omega} k_{zn}^{(3)} \frac{k_{xn}}{k_{n}}\left(e^{i k_{zn}^{(3)} h_2} C_{2n}-e^{-i k_{zn}^{(3)} h_2} C_{2n}^{\prime}\right) \\
\frac{k_{xn}}{k_{n}}\left(e^{i k_{zn}^{(3)} h_2} C_{1n}+e^{-i k_{zn}^{(3)} h_2} C_{1n}^{\prime}\right)+\frac{c}{\sqrt{\varepsilon_{3}} \omega} k_{zn}^{(3)} \frac{k_{y}}{k_{n}}\left(e^{i k_{zn}^{(3)} h_2} C_{2n}-e^{-i k_{zn}^{(3)} h_2} C_{2n}^{\prime}\right)
\end{array}\right)\\
=\left(\begin{array}{c}
-\frac{k_{y}}{k_{n}}\left(T_{1n}+J_{1n}\right)+\frac{c}{\sqrt{\varepsilon_{4}} \omega} k_{zn}^{(4)} \frac{k_{xn}}{k_{n}}\left(T_{2n}-J_{2n}\right) \\
\frac{k_{xn}}{k_{n}}\left(T_{1n}+J_{1n}\right)+\frac{c}{\sqrt{\varepsilon_{4}} \omega} k_{zn}^{(4)} \frac{k_{y}}{k_{n}}\left(T_{2n}-J_{2n}\right)
\end{array}\right)
\end{aligned}
\end{equation}
and 

\begin{equation}
\begin{aligned}
\left(\begin{array}{l}
-\frac{c}{\omega} k_{zn}^{(3)} \frac{k_{xn}}{k_{n}}\left(e^{i k_{zn}^{(3)} h_2} C_{1n}-e^{-i k_{zn}^{(3)} h_2} C_{1n}^{\prime}\right)-\sqrt{\varepsilon_{3}} \frac{k_{y}}{k_{n}}\left(e^{i k_{zn}^{(3)} h_2} C_{2n}+e^{-i k_{zn}^{(3)} h_2} C_{2n}^{\prime}\right) \\
-\frac{c}{\omega} k_{zn}^{(3)} \frac{k_{y}}{k_{n}}\left(e^{i k_{zn}^{(3)} h_2} C_{1n}-e^{-i k_{2n}^{(3)} h_2} C_{1n}^{\prime}\right)+\sqrt{\varepsilon_{3}} \frac{k_{xn}}{k_{n}}\left(e^{i k_{zn}^{(3)} h_2} C_{2n}+e^{-i k_{zn}^{(3)} h_2} C_{2n}^{\prime}\right)
\end{array}\right) \\
=\left(\begin{array}{l}
-\frac{c}{\omega} k_{zn}^{(4)} \frac{k_{xn}}{k_{n}}\left(T_{1n}-J_{1n}\right)-\sqrt{\varepsilon_{4}} \frac{k_{y}}{k_{n}}\left(T_{2n}+J_{2n}\right) \\
-\frac{c}{\omega} k_{zn}^{(4)} \frac{k_{y}}{k_{n}}\left(T_{1n}-J_{1n}\right)+\sqrt{\varepsilon_{4}} \frac{k_{xn}}{k_{n}}\left(T_{2n}+J_{2n}\right)
\end{array}\right).
\end{aligned}
\end{equation}\end{widetext}

\subsection{Scattering matrix}

Finally, we can now write the previous equations in matrix compact form as

\begin{equation}
\begin{aligned}
\left(\begin{array}{l}
\bm{\mathcal{R}} \\
\bm{\mathcal{A}}
\end{array}\right)&=\mathbb{\tilde{S}}_{1}\left(\begin{array}{l}
\bm{\mathcal{I}}\\
\bm{\mathcal{B}}
\end{array}\right),
 \quad\left(\begin{array}{l}
\bm{\mathcal{B}} \\
\bm{\mathcal{C}}
\end{array}\right)=\mathbb{\tilde{S}}_{2}\left(\begin{array}{l}
\bm{\mathcal{A}} \\
\bm{\mathcal{C}}^{\prime}
\end{array}\right),\\
{\rm and} &\quad\left(\begin{array}{l}
\bm{\mathcal{C}^{\prime}} \\
\bm{\mathcal{T}}
\end{array}\right)=\mathbb{\tilde{S}}_{3}\left(\begin{array}{l}
\bm{\mathcal{C}} \\
\bm{\mathcal{J}}
\end{array}\right),
\end{aligned}
\end{equation}
where the explicit expressions of $\mathbb{\tilde{S}}_{1}$, $\mathbb{\tilde{S}}_{2}$, and $\mathbb{\tilde{S}}_{3}$ are already given in section \ref{reflexion_operators}.

Furthermore, to derive the actual scattering matrix, we incorporate a phase correction, as previously mentioned. This adjustment is implemented in Eq. \eqref{smatrix}.

\bibliography{CLP_Graphene_SiO2_Grating.bib}

\end{document}